# Drag or Traction: Understanding How Designers Appropriate Friction in AI Ideation Outputs

A. Baki Kocaballi
University of Technology
Sydney, Australia

Joseph Kizana
University of Technology
Sydney, Australia

Sharon Stein
University of British
Columbia, Canada

Simon Buckingham Shum
University of Technology
Sydney, Australia

## Abstract

Seamless AI presents output as a finished, polished product that users consume rather than shape. This risks design fixation: users anchor on AI suggestions rather than generating their own ideas. We propose *Generative Friction*—intentional disruptions to AI output (fragmentation, delay, ambiguity) designed to transform it from finished product into semi-finished material, inviting human contribution rather than passive acceptance. In a qualitative study with six designers, we identified the different ways in which designers appropriated the different types of friction: users mined keywords from broken text, used delays as workspace for independent thought, and solved metaphors as creative puzzles. However, this transformation was not universal, motivating the concept of *Friction Disposition*—a user's propensity to interpret resistance as invitation rather than obstruction. Grounded in tolerance for ambiguity and pre-existing workflow orientation, Friction Disposition emerged as a potential moderator: high-disposition users treated friction as "liberating," while low-disposition users experienced drag. We contribute the concept of Generative Friction as distinct from Protective Friction, with design implications for AI tools that counter fixation while preserving agency.

## 1 Introduction

Modern generative AI tools are optimized for *seamlessness*—delivering polished, fluent responses instantly. While efficient in some contexts, this presents AI output as a *finished product* to be consumed rather than material to be shaped, risking design fixation [25] and reduced ownership [15].

In design, friction is often treated as drag to be eliminated. But constraints can also be generative: Manning's concept of enabling constraints describes limitations that create conditions for novelty to emerge [20]. Friction in human-AI collaboration can operate similarly—it can be drag (wasted effort) or traction (productive resistance that enables agency).

We propose the concept of **Generative Friction**: intentional disruptions that degrade the seamlessness of AI output to invite human contribution. By fragmenting text (Physical friction), delaying delivery (Temporal friction), or obscuring meaning (Semantic friction), we aim to break the illusion of completeness. While *Protective Friction* adds barriers to verify accuracy in high-stakes tasks [4]—asking "Should I accept this?"—Generative Friction disrupts presentation to ask "What can I *make* from this?" This distinction is critical: creative ideation is low-stakes, where the risk is not accepting a wrong answer but *fixating* on an idea prematurely.

We draw on Dourish's [13] concept of appropriation, the active process of adapting technology to one's own purposes, and Schön's [23] notion of design materials that "talk back." Where seamless AI offers no resistance to prompt reflection, generative friction transforms AI output into material that invites appropriation.

We conducted a qualitative study with six designers using an AI ideation tool under seamless and frictional conditions, asking: *How does intentional friction transform the way designers appropriate AI-generated outputs?* We found that friction transformed the status of AI output from a finished product to accept into semi-finished material requiring further human engagement. Participants *mined* keywords from broken text, *front-loaded* their own ideas during delays, and *translated* metaphors into features. However, this transformation was not universal. We conceptualise **Friction Disposition**—a user's propensity to interpret resistance as invitation rather than obstruction—as a potential moderator. We contribute: (1) the concept of Generative Friction for ideation; (2) empirical evidence of appropriation strategies; and (3) a preliminary model of Friction Disposition.

## 2 Related Work
### 2.1 Friction in HCI and AI

HCI has moved from treating friction as drag to be minimised [21] to recognising its productive potential through Slow Technology's reframing of delay as reflective time [17], Seamful Design's argument that exposing system seams empowers users [7], and Cox et al.'s operationalisation of "microboundaries" for mindfulness [10]. In the AI domain, this lineage continues with Cabitza et al.'s programmed inefficiencies that stimulate cognitive engagement [6], and Chen & Schmidt's [8] behavioral model of Positive Friction. We build on this trajectory, introducing Generative Friction as friction designed to degrade AI output seamlessness and stimulate user contribution.

## 2.2 Design Fixation and Appropriation

Seamless AI risks automation complacency [22] and design fixation: users anchor on AI-generated ideas, constraining divergent thinking. Wadinambiarachchi et al. [25] found that AI during ideation increased fixation and proposed "partially completed or blurred outputs" as countermeasure. Prior work supports this direction: ambiguous, incomplete or misaligned AI outputs can work as generative resources [12], and partial photographs of products can reduce design fixation [9]. Cognitive Forcing Functions [4] interrupt fast, intuitive acceptance (System 1) and trigger deliberate evaluation (System 2) [18]. Seamful XAI [14] extends this to explainability, arguing that revealing AI seams increases agency.

However, this work predominantly focuses on high-stakes domains such as aviation, medicine, decision-making, where the goal is preventing costly errors. Creative ideation is *low-stakes*: the risk is not accepting a wrong answer, but *fixating* on an idea prematurely without exploring a larger space of possibilities. Friction in ideation should therefore stimulate elaboration rather than verify accuracy.

As introduced earlier, Dourish's [13] appropriation framework and Schön's [23] concept of reflective resistance provide our theoretical lens. Where seamless AI presents output as finished—leaving little "back-talk"—our study examines how generative friction increases appropriability by inviting users to work *with* the output rather than accept it.

## 3 Method

We conducted a qualitative, within-subjects study using think-aloud protocols to compare unrestricted (seamless) and restricted (friction) AI-assisted ideation tasks. Each session began with a short pre-task interview, followed by the ideation tasks. Low risk ethics approval was received, and sessions lasted 45–60 minutes.

### 3.1 Participants

We recruited 6 participants (3 female, 3 male; ages 22–30) with design backgrounds (Table 1). Three were students studying Master of Design and three were working professionals (Table 2).

Table 1: Participant Demographics

| ID | Age/Gender | Role | GenAI Use |
|---|---|---|---|
| P1 | 22M | Postgrad Student | Occasional |
| P2 | 23M | Postgrad Student | Occasional |
| P3 | 23F | Postgrad Student | Regular |
| P4 | 24F | Product designer (2 yrs) | Regular |
| P5 | 27F | Product designer (5 yrs) | Regular |
| P6 | 30M | UX/UI designer (7 yrs) | Regular |

Because our friction conditions deliberately introduce ambiguity into AI output, we anticipated that participants' tolerance for ambiguity—their capacity to engage productively with uncertain or incomplete information [5], a trait positively associated with creative behaviour [26]—could influence their responses. We therefore assessed each participant's orientation toward AI output and ambiguity qualitatively, based on pre-task interview statements prior to any exposure to friction conditions. Future work will complement this qualitative coding with a standardised ambiguity tolerance scale (Table 2).

Table 2: Participant Predispositions toward AI and Ambiguity

| ID | Orientation | Representative Quote |
|---|---|---|
| P4 | Low Tolerance | "I just don't see the point of staring at a blank page anymore." |
| P6 | Low Tolerance | "I use it when I just really can't be bothered." |
| P1 | High Tolerance | "It shouldn't do everything for you." |
| P2 | High Tolerance | "There's nothing challenging your thinking [with AI]." |
| P3 | High Tolerance | "I avoid using it before I've sketched something myself." |
| P5 | High Tolerance | "Creativity comes from thinking outside the box." |

### 3.2 System and Conditions

We developed **SPARK v1**, a custom AI ideation tool (GPT-4o). The underlying model and system prompt were constant; the *presentation* of output varied across four conditions:

(1) **Seamless (Baseline):** Complete AI output displayed normally without any intervention.
(2) **Physical Friction (Fragmentation):** AI Output visually broken— every second word of the full idea text is hidden, causing fragmented text blocks. Designed to the **disfluency effect** [1].
(3) **Temporal Friction (Delay):** The words of the full AI output are displayed one by one gradually over time. Designed to the **incubation effect** [24].
(4) **Semantic Friction (Ambiguity):** AI Output was cryptic involving a metaphorical description of the idea. Designed to **abstract stimuli** [16] and **semantic ambiguity** [12]

### 3.3 Procedure and Analysis

The study consisted of pre-task interviews (assessing AI experience and attitudes), four 7-minute ideation tasks responding to design briefs, and post-task interviews. The four briefs were selected to minimise variation while maintaining a similar cognitive task profile: all were student-support product design problems in a shared domain: designing a product, app, or service to help students (1) manage productivity, (2) collaborate as a group, (3) plan for the future, and (4) reflect on what they had learnt, requiring comparable levels of problem framing, constraint satisfaction, and feature ideation. We set the ideation window to 7 minutes, informed by our pilot study, to provide sufficient time for concept generation while maintaining task focus. Conditions were presented in fixed order (Seamless → Physical → Temporal → Semantic). We acknowledge this confounds friction type with order and fatigue effects; we return to this limitation in Section 6. We used a process-oriented, reflexive thematic analysis on transcribed data, focusing on "breakdowns" (moments where friction halted progress) and "repair strategies" (how users overcame the halt) to surface specific appropriation moves [2].

## 4 Findings

We found that intentional friction did not simply "break" the user's workflow; participants actively appropriated the disruption, transforming obstacles into creative tools. However, this appropriation appeared to have depended on the user's underlying orientation toward ambiguity and constraint.

### 4.1 Strategies of Appropriation

Participants developed three primary strategies to appropriate degraded AI output.

**Keyword Mining (Physical Friction).** When fragmented output prevented sentence-level reading, participants shifted to **keyword mining**—scanning broken text for actionable nouns and verbs. P3 described this as *accelerated reading*: *"It was pulling out keywords for me. So I didn't have to go through the text and highlight"* (P3). The broken text became a "tag cloud" for immediate concept extraction. P6 similarly noted: *"I'm seeing bits and pieces now. It's just triggering a different thought."* Notably, P3's case was distinctive: keyword mining was their pre-existing AI workflow practice, meaning physical fragmentation was functionally invisible as friction—a point we return to in Section 4.3.

**Parallel Processing & Front-Loading (Temporal Friction).** Under delays, participants filled the void with independent ideation rather than waiting passively. P5 described **front-loading**: *"I went to there [my own mind], came up with a few ideas while I was waiting... and then you can come back and tweak"* (P5). This shifted the usual dynamic: participants reported leading with their own ideas before consulting AI output, though we acknowledge the study design cannot fully distinguish deliberate front-loading from idle time-filling during the enforced wait. P2 similarly noted: *"I like to think I had some good ideas in the time that it took."*

**Abstract Interpretation (Semantic Friction).** Semantic friction, the cryptic, metaphorical outputs, forced **interpretive labor**. Unlike physical friction (which obscured form) or temporal friction (which delayed delivery), semantic friction obscured *meaning itself*. For ambiguity-tolerant participants, this became puzzle-solving: *"I feel like you've got given a different medium, like jelly. But it allowed you to move more fluid within it"* (P5). P1 described the experience as a "mini-game," actively seeking connections between abstract metaphors and concrete design problems: *"Thinking of a labyrinth makes me think of like a maze... so maybe a game app."*

### 4.2 "Friction Disposition"

Appropriation and appreciation of friction were not universal; it depended on users' tolerance for ambiguity—their readiness to engage with uncertain, incomplete, or metaphorical content. High-tolerance users treated friction as invitation; low-tolerance users experienced it as obstruction (Table 3).

We propose that Friction Disposition relates to at least two measurable factors: (a) *tolerance for ambiguity*—a well-studied personality construct [5]—and (b) *pre-existing workflow orientation* toward AI output. P3, who already parsed AI text by keywords rather than sentences, experienced physical friction as invisible. P4, who relied on copy-paste workflows optimised for speed, experienced all friction as obstruction. In our data, these pre-existing orientations appeared to align with friction responses at least as strongly as friction type alone.

### 4.3 Preliminary Dispositional Profiles

Though preliminary, our data with six designers reveals three distinct archetypes of what we term Friction Disposition.

**The Reframers** (P1, P5) interpreted friction as creative constraint. P1 exhibited playful engagement—treating friction as a puzzle: *"I could genuinely feel my head throbbing to think... usually when I'm using AI I'm not thinking."* P5 described semantic friction as liberating: *"I could go wherever I wanted, like a constellation, a forest. It didn't give me the answer, so I had to make it"* (P5).

**The Fluent Appropriator** (P3) represents users for whom friction becomes *invisible*. As noted in Section 4.1, P3's baseline keyword-processing practice meant physical fragmentation was not experienced as friction at all. Furthermore, P3 treated semantic metaphors not as riddles but as "role reversal": *"It's forcing me to have to use my own brain... It's almost like a role reversal"* while producing the highest idea count under this condition Table 4.

**The Resisters** (P4, P6) perceived friction as obstruction. P4: *"I don't like it at all. It's just annoying. I want it to give me the ideas so I can move on"* (P4). P6 exhibited friction fatigue: they adapted to physical and temporal friction early but collapsed under semantic friction: *"This is where I would tap out... This hurts my brain to try to interpret."* (We note that semantic friction was always the final condition, so cumulative fatigue may have contributed to P6's collapse; see Section 6.)

### 4.4 A Dispositional Contrast: P4 vs P3

The P3–P4 contrast most clearly illustrates Friction Disposition as a moderator. Both experienced identical conditions yet diverged dramatically. P3 entered the study with a keyword-processing mental model (*"I never copy-paste whole ideas"*), and this pre-existing orientation meant friction was absorbed into workflow rather than fought against—physical fragmentation was frictionless, temporal delays became background processing time, and semantic ambiguity produced the highest engagement. P4, by contrast, prioritised speed and efficiency (*"efficiency is a big part of my process"*), experiencing each friction type as obstruction to be endured or rejected. This divergence under identical experimental conditions suggests that friction appropriation is *relational*, depending not only the friction design but also on the user's pre-existing readiness to engage with resistance—their Friction Disposition (see also Table 3).

## 5 Discussion

Our findings extend the discourse on friction in human-AI collaboration for AI-assisted creative ideation, a low-stakes domain where the cost of a "bad" idea is tolerable compared to a high-stakes domain with strict requirements of safety and reliability. We see three paradoxes playing out, which merit further investigation:

**Friction: Drag or Traction?** In physics, friction provides *traction*—the grip that enables controlled motion. Our findings reveal the same duality. P1's disposition converted friction into **traction**:

Table 3: Participant appropriation trajectories and strategies across friction conditions.

| ID | Archetype | Physical Friction | Temporal Friction | Semantic Friction |
| --- | --- | --- | --- | --- |
| P1 | Reframer | **Breakdown→Repair:** decoding/puzzle solving. | **Breakdown→Repair:** front-loading (sketching while waiting). | **Breakdown→Repair:** abstract interpretation ("mini-game"). |
| P2 | Reframer | **Breakdown→Repair:** keyword mining/extraction. | **Breakdown→Repair:** anticipatory ideation during delay. | **Endured/Worked-around:** mixed uptake; limited interpretation. |
| P3 | Fluent Appropriator | **Absorbed:** keyword mining as baseline practice. | **Absorbed:** parallel processing; AI queued in background. | **Breakdown→Repair:** abstract interpretation ("role reversal"). |
| P4 | Resister | **Rejected:** abandoned interpretation. | **Rejected:** impatient waiting/bypassing. | **Rejected:** re-rolled for concreteness. |
| P5 | Reframer | **Breakdown→Repair:** keyword mining (slower but functional). | **Breakdown→Repair:** front-loading own ideas first. | **Breakdown→Repair:** abstract interpretation ("like jelly"). |
| P6 | Resister | **Rejected:** incomplete ideas unhelpful. | **Rejected:** delays as failure. | **Collapse:** tapped out under interpretive load. |

productive resistance that enabled agency. P4's disposition experienced only **drag**: wasted effort that blocked workflow. The same design produced opposite valences depending on disposition.

**The Seamlessness Trap.** A central tension emerged: users who *most* need friction (over-trusters like P4) are *least* likely to accept it. P4 copy-pasted AI output verbatim in the seamless condition—demonstrating precisely the over-reliance that friction targets—yet rejected friction outright. Buçinca et al. [4] found effective interventions received the worst ratings. Unlike their uniform finding, we observed variance: P1 called friction "like a game"; P5 found it "liberating." This suggests the effectiveness-acceptance trade-off depends on *who* the user is. This pattern resonates with psychological reactance theory: users who perceive friction as threatening their efficiency, a valued behavioral freedom, resist the intervention most strongly, regardless of its potential benefit [3].

**Friction Fatigue.** P6's collapse in the final condition suggests friction tolerance may be a function of **cumulative cognitive load**. Sustained interpretive effort under friction depletes attentional resources, leading to breakdown—a framing grounded in cognitive load theory rather than the contested ego depletion model. Future designs should consider fatigue-aware adaptation by reducing friction intensity as sessions progress or cognitive load exceeds a threshold. Prior work on friction has focused on *whether* to introduce friction, not *who bears the cost*. However, friction could be shareable: future designs might distribute the burden through AI self-interpretation (offering multiple readings of its own metaphor) or progressive disclosure (cryptic output with an "explain" toggle). Table 5 in Appendix shows the improved version of Spark implementing tuneable frictions.

**Design Implications.** Based on our findings, we propose four principles for designing Generative Friction:

- *Mode-selectable:* Offer exploration, efficiency, and validation modes rather than a single intensity dial. As P3 noted, some users already "parse text by keywords," making Physical friction invisible for some and obstructive for others.
- *Legible:* Explicitly communicate friction's purpose. Instead of "You may experience incomplete suggestions," try "This tool shows fragments to help you build ideas, not copy them"—reframing friction from "broken feature" to "designed affordance."
- *Burden-shared:* Distribute cognitive work between user and AI through mechanisms like progressive disclosure or AI self-interpretation.
- *Escapable:* Preserve agency through opt-out mechanisms, since inescapable friction drives abandonment [19]. Friction should be a *nudge*, not a *trap*.

We note a productive tension between legibility and escapability: making friction's purpose visible may reduce the need for escape, while easy escape may undermine the generative intent. Future work should explore how these principles interact in practice. We note also that Friction Disposition's focus on readiness to engage with ambiguity and uncertainty resonates directly with professional and student "Learning Dispositions" [11] that slows learners down to reflect on deep-seated assumptions. We see a key opportunity to investigate friction design for deeper learning.

Our sample (N=6) allowed rich qualitative insight but limits generalizability. Conditions were presented in fixed order (Seamless → Physical → Temporal → Semantic), so we cannot fully disentangle friction effects from order effects or fatigue. Future work should counterbalance conditions, develop a validated Friction Disposition scale, and explore adaptive friction systems that respond to user state in real time.

## 6 Conclusion

We investigated whether intentional friction can transform AI output from finished product into semi-finished material requiring further human engagement. We found that friction enabled different appropriation methods including disappropriation. Our findings suggest user appropriation is moderated by Friction Disposition. High-disposition users described friction as "liberating" and "like a game"; low-disposition users experienced only drag. We propose two conceptual contributions: the concept of Generative Friction to stimulate human engagement and Friction Disposition as a potential moderating factor.

## A Appendix: Idea Counts across Sessions and SPARK v1 vs SPARK v2 (With Tuneable Friction)

While Table 4 provides a summary of the idea counts across all the sessions, Table 5 summarises the key differences between SPARK v1 (used in the current study) and the redesigned SPARK v2, which incorporates tuneable friction controls informed by our findings. SPARK v2's design directly implements two of our proposed design principles: *Mode-selectable* (users choose their friction level) and *Escapable* (progressive disclosure allows opting out of high friction without abandoning the tool entirely). The tuneable controls also operationalise the *Burden-shared* principle by distributing interpretive effort between the system's default friction and the user's chosen level of engagement.

**Table 4: Idea counts per participant across seamless and friction conditions.**

| Participant | Seamless | Physical | Temporal | Semantic |
|---|---|---|---|---|
| P1 | 6 | 5 | 6 | 6 |
| P2 | 5 | 1 | 1 | 1 |
| P3 | 9 | 7 | 11 | 14 |
| P4 | 5 | 3 | 5 | 4 |
| P5 | 4 | 3 | 5 | 5 |
| P6 | 5 | 4 | 5 | 2 |

Table 5: Comparison of SPARK v1 and SPARK v2 friction conditions. SPARK v2 introduces user-controllable mechanisms that allow participants to modulate friction intensity, addressing the Friction Disposition differences observed in the study.

| Friction Type | SPARK v1 | SPARK v2 (Tuneable) |
|---|---|---|
| Physical | 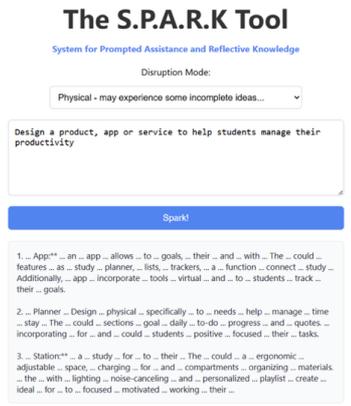 Every second words are hidden, creating a fragmented presentation with ellipses between words. | 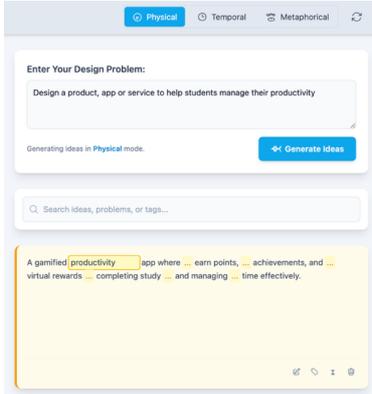 Fragmented display is retained, but users can **click on the ellipsis** to reveal hidden words, progressively reducing friction on demand. |
| Temporal | 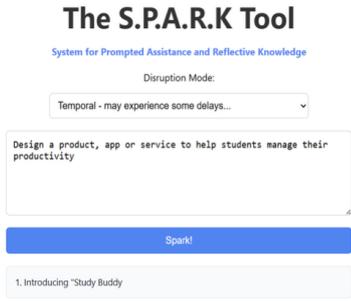 AI output text is displayed gradually over time. | 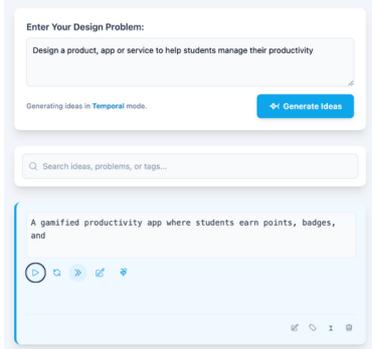 Users can **play, pause, and speed up** the presentation of ideas using playback controls, giving them agency over pacing. |
| Semantic | 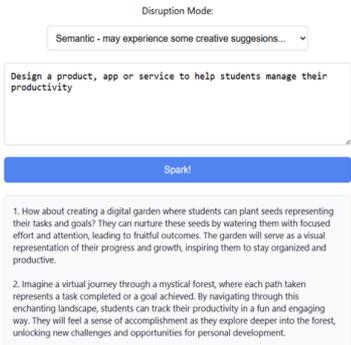 Output is presented as cryptic metaphors and riddle-like text. Users must interpret abstract language to extract design ideas. | 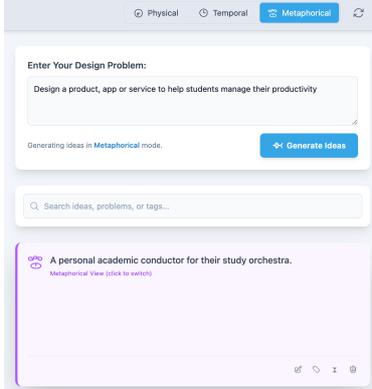 Metaphorical output is retained, but users can **click on the idea box** to reveal a more literal description, providing an "explain" toggle. |